\documentclass[aps,prb,twocolumn,showpacs,psfig,superscriptaddress,longbibliography]{revtex4-1}
\usepackage{amsfonts}
\usepackage{mathrsfs}
\usepackage{amsmath}
\usepackage{color}
\usepackage{natbib}
\usepackage{textcomp}
\usepackage{graphicx}
\usepackage{bm}
\usepackage{amssymb}

\usepackage{xspace}
\usepackage{epstopdf}
\usepackage{dcolumn}
\usepackage{longtable}
\usepackage{multirow}
\usepackage[colorlinks=true, letterpaper=true, pdfstartview=FitV, linkcolor=blue, citecolor=blue, urlcolor=blue]{hyperref}
\usepackage{float}

\makeatletter

\newcommand{\Rmnum}[1]{\expandafter\@slowromancap\romannumeral #1@}
\makeatother

\begin{document}

\title{Electric field induced topological phase transition and large enhancements of spin-orbit coupling and Curie temperature in two-dimensional ferromagnetic semiconductors}

\author{Jing-Yang You}
\affiliation{Kavli Institute for Theoretical Sciences, and CAS Center for Excellence in Topological Quantum Computation, University of Chinese Academy of Sciences, Beijing 100190, China}

 \author{Xue-Juan Dong}
 \affiliation{School of Physical Sciences, University of Chinese Academy of Sciences, Beijing 100049, China}

\author{Bo Gu}
 \email{gubo@ucas.ac.cn}
 \affiliation{Kavli Institute for Theoretical Sciences, and CAS Center for Excellence in Topological Quantum Computation, University of Chinese Academy of Sciences, Beijing 100190, China}
\affiliation{Physical Science Laboratory, Huairou National Comprehensive Science Center, Beijing 101400, China}
 
\author{Gang Su}
\email{gsu@ucas.ac.cn}
\affiliation{Kavli Institute for Theoretical Sciences, and CAS Center for Excellence in Topological Quantum Computation, University of Chinese Academy of Sciences, Beijing 100190, China}
\affiliation{School of Physical Sciences, University of Chinese Academy of Sciences, Beijing 100049, China}
\affiliation{Physical Science Laboratory, Huairou National Comprehensive Science Center, Beijing 101400, China}

\begin{abstract} 
Tuning topological and magnetic properties of materials by applying an electric field is widely used in spintronics. In this work, we find a topological phase transition from topologically trivial to nontrivial states at an external electric field of about 0.1 V/\AA\ in MnBi$_2$Te$_4$ monolayer that is a topologically trivial ferromagnetic semiconductor. It is shown that when electric field increases from 0 to 0.15 V/\AA, the magnetic anisotropy energy (MAE) increases from about  0.1 to 6.3 meV, and the Curie temperature Tc increases from 13 to about 61 K. The increased MAE mainly comes from the enhanced spin-orbit coupling due to the applied electric field. The enhanced Tc can be understood from the enhanced $p$-$d$ hybridization and decreased energy difference between $p$ orbitals of Te atoms and $d$ orbitals of Mn atoms. Moreover, we propose two novel Janus materials MnBi$_2$Se$_2$Te$_2$ and MnBi$_2$S$_2$Te$_2$ monolayers with different internal electric polarizations, which can realize quantum anomalous Hall effect (QAHE) with Chern numbers $C$=1 and $C$=2, respectively. Our study not only exposes the electric field induced exotic properties of MnBi$_2$Te$_4$ monolayer, but also proposes novel materials to realize QAHE in ferromagnetic Janus semiconductors with electric polarization.
\end{abstract}
\maketitle

\section{Introduction}
Quantum anomalous Hall effect (QAHE) as a novel topological phase has attracted tremendous interest because of its potential applications in dissipationless spintronics~\cite{Haldane1988,Onoda2003,Liu2008,Wu2008,Yu2010,Chang2013,Wu2014,Mogi2015,Chang2015,Liu2016,Si2017,You2019b,Ou2017,He2018,You2019,Li2020}. The QAHE was firstly realized experimentally in Cr-doped (Bi, Sb)$_2$Te$_3$ thin film at 30 mK~\cite{Chang2013} and later in V-doped (Bi, Sb)$_2$Te$_3$ thin film at 25 mK~\cite{Chang2015} and Cr-doped (Bi, Sb)$_2$Te$_3$ thin film at about 2 K~\cite{Mogi2015}. Then, the QAHE was observed in a Cr- and V-codoped (Bi, Sb)$_2$Te$_3$ system at about 300 mK~\cite{Ou2017}. Since magnetic order that is stable at high temperature is essential for QAHE, we need to increase the ordering temperature of a magnetically-doped topological insulator.

MnBi$_2$Te$_4$ as a new platform to realize QAHE has gained extensive studies both in theory and in experiments~\cite{Otrokov2017,Otrokov2019a}. MnBi$_2$Te$_4$ is composed of septuple Te–Bi–Te–Mn–Te–Bi–Te sequences, and exhibits a van der Waals layered structure~\cite{Lee2013}. Bulk MnBi$_2$Te$_4$ is an antiferromagnetic insulator with the Ne$\acute{e}$l temperature of 25 K~\cite{Cui2019,Li2020a}, which presents the axion state~\cite{Zhang2019}. Topological surface states with diminished gap forming a characteristic Dirac cone attributed to multidomains of different magnetization orientations were observed in antiferromagnetic topological insulator MnBi$_2$Te$_4$~\cite{Chen2019}. The heterostructures (Bi$_2$Te$_3$)$_n$(MnBi$_2$Te$_4$) were extensively studied to realize QAHE.~\cite{Vidal2019,Aliev2019,Deng2020a,Lei2020,Rienks2019,Hu2020,Klimovskikh2020}. MnBi$_2$Te$_4$ monolayer is a topologically trivial ferromagnetic semiconductor~\cite{Li2019a,Li2019,Otrokov2019}, while its multilayers host the states alternating between QAH and zero plateau QAH for odd and even number of monolayers, respectively~\cite{Otrokov2019}. In a five-septuple-layer MnBi$_2$Te$_4$, the QAHE was observed at 1.4 K, and the quantization temperature can be raised up to 6.5 K by an external magnetic field to align all layers ferromagnetically~\cite{Deng2020}. High-Chern-number QHE without Landau levels was obtained in ten-septuple-layer MnBi$_2$Te$_4$ under applied magnetic field and back gate voltages~\cite{Ge2020}. In six-septuple-layer MnBi$_2$Te$_4$, the axion insulator state occurs over a wide magnetic field range and at relatively high temperatures, while a moderate magnetic field can drive the axion insulator phase to a Chern insulator phase with QAHE~\cite{Liu2020}.

Tuning the energy gap is widely used in spintronics, because it can optimize the properties of materials and make materials have better performance in devices. There are several ways to tune the energy bands, such as by an electric field~\cite{Weisheit2007,Deng2018,Jiang2018,Du2020}, strain~\cite{Ni2008,Liu2013,Dong2019}, doping~\cite{Sim2015,You2019a,Ahmed2015,You2020}, vacancy~\cite{Chen2017,Liu2019,Hou2020}, surface modification~\cite{Sim2015,Nemani2018} and so on. Tuning energy bands by an external electric field has some advantages: the magnitude and direction of the external electric field can be arbitrarily controlled; applying the external electric field is feasible and easy to realize in experiments. Graphene is a two-dimensional (2D) material with a zero band gap that restricts its application in electronic devices. After applying a gate voltage, the inversion symmetry is broken and a nonzero band gap is opened~\cite{Ohta2006,Min2007,Zhang2009,Mak2009,Lui2011}. The band engineering by external electric fields was also performed in MoS$_2$~\cite{Lu2014}. 

In this work, we study novel properties of MnBi$_2$Te$_4$ monolayer controlled by an electric field. With the increase of electric field, the band gap decreases and the band inversion with a topological phase transition occurs at an electric field of about 0.1 V/\AA, and the system then enters into a topological state. With the electric field ranging from 0 to 0.15 V/\AA, the magnetic anisotropy energy (MAE) increases from about 0.1 to 6.3 meV, and the Curie temperature Tc increases from 13 to about 61 K. The increased MAE mainly comes from the enhanced spin-orbit coupling (SOC) due to the applied electric field. By means of the super-exchange picture, the enhanced Tc can be understood from the enhanced $p$-$d$ hybridization and decreased energy difference between $p$ orbitals of Te atoms and $d$ orbitals of Mn atoms. Inspired by the electric field induced topological phase transition, we propose 2D Janus materials MnBi$_2$Se$_2$Te$_2$ and MnBi$_2$S$_2$Te$_2$ with different internal electric polarizations, which can realize QAHE. Our findings not only reveal the electric field induced exotic properties of MnBi$_2$Te$_4$ monolayer, but also present a new idea to realize QAHE in novel ferromagnetic Janus materials with electric polarizations.

\section{Calculation method}
Our first-principles calculations were based on density functional theory (DFT) as implemented in the Vienna Ab initio Simulation Package (VASP)~\cite{Kresse1996}, using the projector augmented-wave method~\cite{Bloechl1994}. The generalized gradient approximation with the Perdew-Burke-Ernzerhof~\cite{Perdew1996} realization was adopted for the exchange-correlation functional. We set the on-site effective Hubbard interaction $U$ = 4 eV in considering electron correlation of 3$d$ electrons of Mn atoms. The monolayer was placed under a vacuum layer of 15 \AA. The plane-wave cutoff energy was set to 550 eV. A Monkhorst-Pack k-point mesh~\cite{Monkhorst1976} with a size of 19$\times$19$\times$1 was used for the BZ sampling. The crystal structure was optimized until the forces on the ions were less than 0.0001 eV/\AA, and the total energy was converged to 10$^{-8}$ eV with Gaussian smearing method. The interaction of the magnetic dipoles was taken into account in calculations of the magnetic anisotropy energies. The phonon spectra were obtained in a 4$\times$4$\times$1 supercell with the PHONOPY package~\cite{Togo2015}. The surface spectrum was calculated by using the Wannier functions and the iterative Green's function method~\cite{Marzari1997, Souza2001, Wu2018a, Sancho1985}.

\begin{figure}[!!htbp]
	\includegraphics[width=8.5cm]{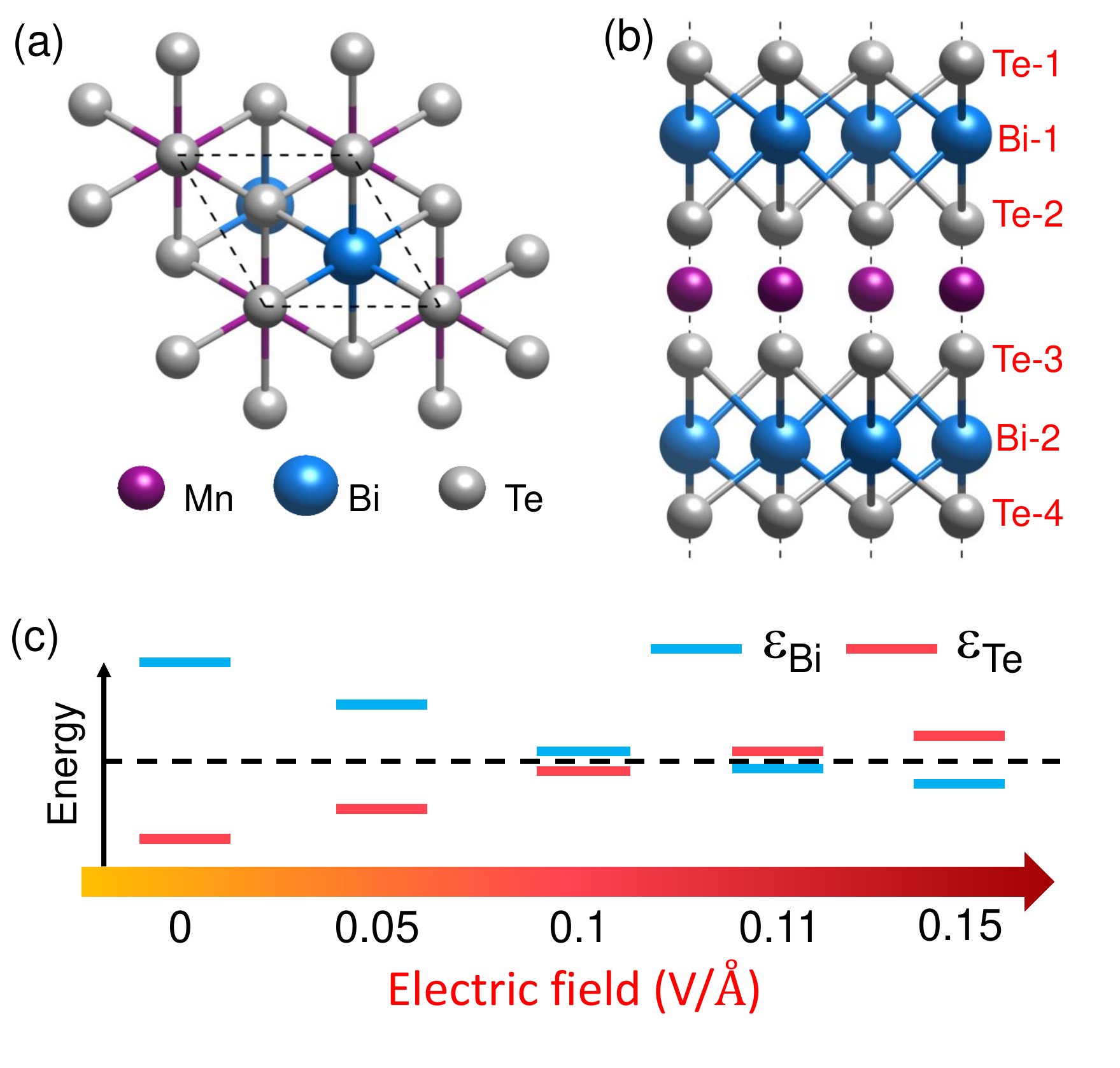}
	\caption{Crystal structure of monolayer MnBi$_2$Te$_4$. The (a) top and (b) side view of monolayer MnBi$_2$Te$_4$, where the atoms are labeled with elements and atomic layer number in (b). (c) The evolution of the top of valence band and the bottom of conduction band relative to their energy center at $\Gamma$ point with increasing the electric field.}
	\label{F-1}
\end{figure}

\section{Monolayer MnBi$_2$Te$_4$}
\begin{figure}[!!!htbp]
 	\includegraphics[width=8.5cm]{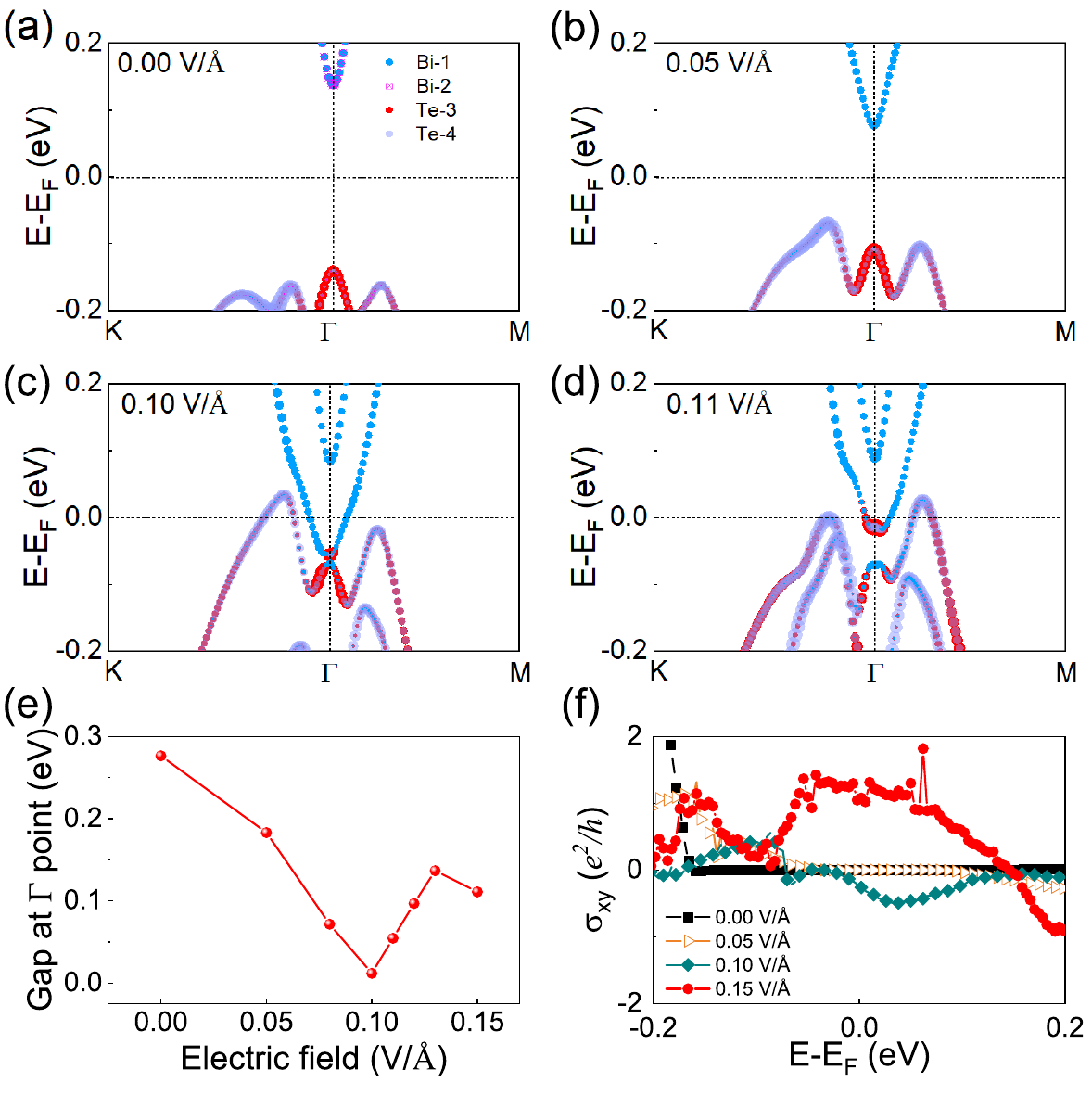}
 	\caption{Electric field induced topological phase transition. Atom projected electronic band structures of monolayer MnBi$_2$Te$_4$ under the electric field of (a) 0.00, (b) 0.05, (c) 0.10 and (d) 0.11 V/\AA, respectively. The electric field dependent (e) direct band gap at $\Gamma$ point and (f) anomalous Hall conductivity (AHC). The results were obtained by the GGA+SOC+$U$ calculations with $U$ = 4 eV.}
 	\label{F-2}
 \end{figure}

Monolayer MnBi$_2$Te$_4$ can be grown in a quintuple layer of Bi$_2$Te$_3$ and a bilayer of MnTe with the molecular beam epitaxy method~\cite{Gong2019}. The crystal structure of monolayer MnBi$_2$Te$_4$ is shown in Fig \ref{F-1}. The space group is P$\bar{3}$m1 (No. 164) with an optimized lattice constant $a$ = 4.34 \AA. Monolayer MnBi$_2$Te$_4$ is a ferromagnetic semiconductor with out-of-plane magnetization, and the Curie temperature T$_C$ of its seven septuple-layers film is about 20 K~\cite{Gong2019}. Our estimated Curie temperature of monolayer MnBi$_2$Te$_4$ is comparable with that of its film~\cite{Gong2019,SuplMat}. The atom projected band structure of monolayer MnBi$_2$Te$_4$ without electric field was plotted in Fig. \ref{F-2}(a), from which one can observe that the bottom of conduction band of monolayer MnBi$_2$Te$_4$ is mainly attributed to Bi-1 and Bi-2 atoms, while the main contribution to the top of valence bands is from Te-4 and Te-1 atoms. Because of the inversion symmetry, Bi-1 and Bi-2 atoms give equal contributions to the density of states of the bottom of conduction bands at $\Gamma$ point, and Te-1 and Te-4 atoms contribute equally to the density of states of the top of valence bands at $\Gamma$ point.
 
\section{Monolayer MnBi$_2$Te$_4$ under an electric field}
By applying an electric field E normal to the atoms plane, we can tune the properties of monolayer MnBi$_2$Te$_4$. The effect of electric field on the properties of monolayer MnBi$_2$Te$_4$ is summarized in Table~S1. There are three main changes of the properties of monolayer MnBi$_2$Te$_4$ with increasing the electric field, including the changes of band gap, magnetic anisotropy and exchange integral.  
 
\subsection{Evolution of band gap under an electric field}
The band structure of monolayer MnBi$_2$Te$_4$ varies with increasing the electric field without a big deformation of profiles but with a remarkable change of the band gap as shown in Fig. \ref{F-2}. The evolution of the band gap at $\Gamma$ point with the electric field was schematically plotted in Fig. \ref{F-1} (c). The direct band gap at $\Gamma$ point [Fig.~\ref{F-2}(e)] as well as the global band gap increases with increasing the electric field lower than 0.1 V/\AA. This is intuitive, because on one hand, the electric field E breaks the inversion symmetry of monolayer MnBi$_2$Te$_4$ resulting in the degeneracy broken of Bi-1 and Bi-2, Te-1 and Te-4 atoms at $\Gamma$ point, respectively, and on the other hand, the effect of electric field on the upper triple Te-Bi-Te layers consisting of Te-1, Bi-1, and Te-2 atoms is equivalent to electron doping, while for the lower triple Te-Bi-Te layers consisting of Te-3, Bi-2, and Te-4 atoms it corresponds to hole doping. Thus, for the cation Bi-1, its energy level moves downward close to Fermi level, while for anion Te-4, its energy level moves upward close to Fermi level. This analysis is also verified by our DFT results as shown in Fig. \ref{F-2}. Therefore, the energy levels of Bi-1 and Te-4 atoms should be touched at a proper electric field, which is about 0.1 V/\AA\ in our calculations. When the electric field continues to increase to 0.11 V/\AA, an inversion of the energy bands occurs at $\Gamma$ point [Fig.~\ref{F-2}(d)] with a topological phase transition from a topological trivial state to a nontrivial state with nonzero Chern number $C$=1. However, because there is no global energy gap, we cannot observe a quantized AHC, but it can adiabatically evolve to the quantum anomalous Hall insulating state. When the electric field continues to increase from 0.13 to 0.15 V/\AA, the direct band gap at $\Gamma$ point decreases, and the system turns into the topological state with $C$=2. Although there is also no global gap, it is interesting to note that a relatively flat plateau of AHC occurs with E=0.15 V/\AA\ near Fermi level as shown in Fig.~\ref{F-2}(f).
 
\begin{figure}[!!!!htbp]
	\includegraphics[width=8.5cm]{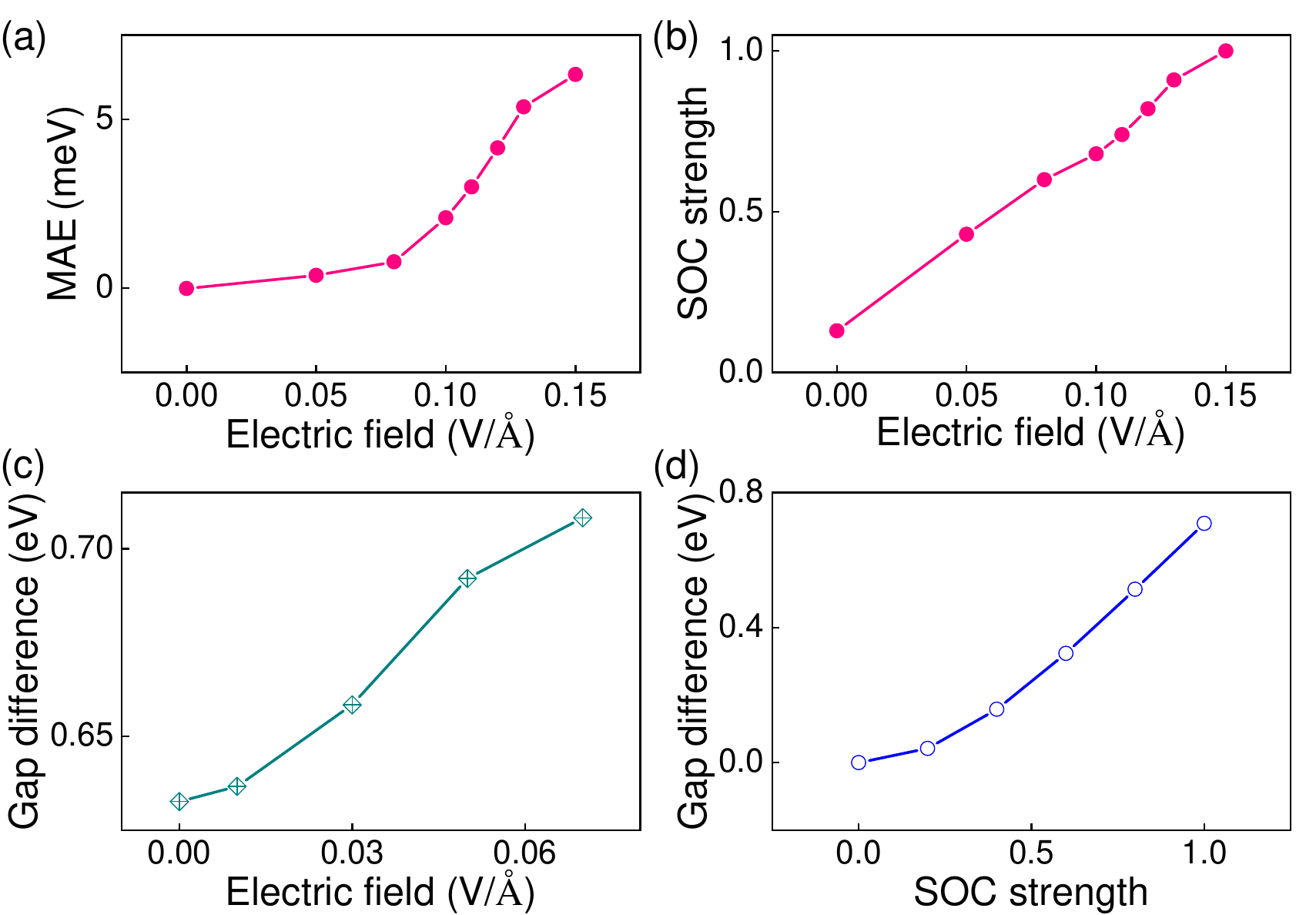}
	\caption{Electric field enhanced spin-orbit coupling. The (a) electric field and (b) SOC strength dependent MAE, where MAE means the energy difference per MnBi$_2$Te$_4$ formula unit between the in-plane and out-of-plane magnetic configurations and the SOC strength under the electric field of 0.15 V/\AA\ is set to 1. The (c) electric field and (d) SOC strength dependent gap difference between the gaps without and with SOC.}
	\label{F-3}
\end{figure}

\subsection{Magnetic anisotropy of MnBi$_2$Te$_4$ monolayer under an electric field}
Besides the topological phase transition, the electric field also dramatically influences magnetic anisotropy energy (MAE), where MAE means the energy difference per MnBi$_2$Te$_4$ formula unit between the out-of plane and in-plane ferromagnetic configurations. The electric field E dependent MAE was plotted in Fig.~\ref{F-3}(a). From Fig.~\ref{F-3}(a), one may observe that MAE increase with increasing E. MAE are usually related to the SOC, thus the increase of MAE may come from the enhanced SOC because of the electric field. To verify this observation, we extracted the dependence of MAE on the SOC strength with the electric field of 0.15 V/\AA\ from DFT calculations,  and obtained the electric field dependent SOC strength by comparing the MAE under different electric fields and SOC strength as plotted in Fig. \ref{F-3}(b). It is obvious that the SOC is enhanced with increasing the electric field. A larger anomalous Hall conductivity (AHC) $\sigma_{xy}$ is expected in monolayer MnBi$_2$Te$_4$ with enhanced SOC. We calculated the AHC with electric fields of 0.05, 0.10 and 0.15 V/\AA, respectively, as shown in Fig.~\ref{F-2}(f). Enhanced SOC can also be seen from the enhanced AHC as shown in Fig.~\ref{F-2}(f) and the gap difference between the band gaps without SOC and with SOC in Figs.~\ref{F-3}(c) and (d) with increasing the electric field. The Mn local magnetic moment is not sensitive to the electric field, and in the range of electric field we studied, it keeps at about 4.5 $\mu_B$.

\begin{figure}[!!!htbp]
\includegraphics[width=8.7cm]{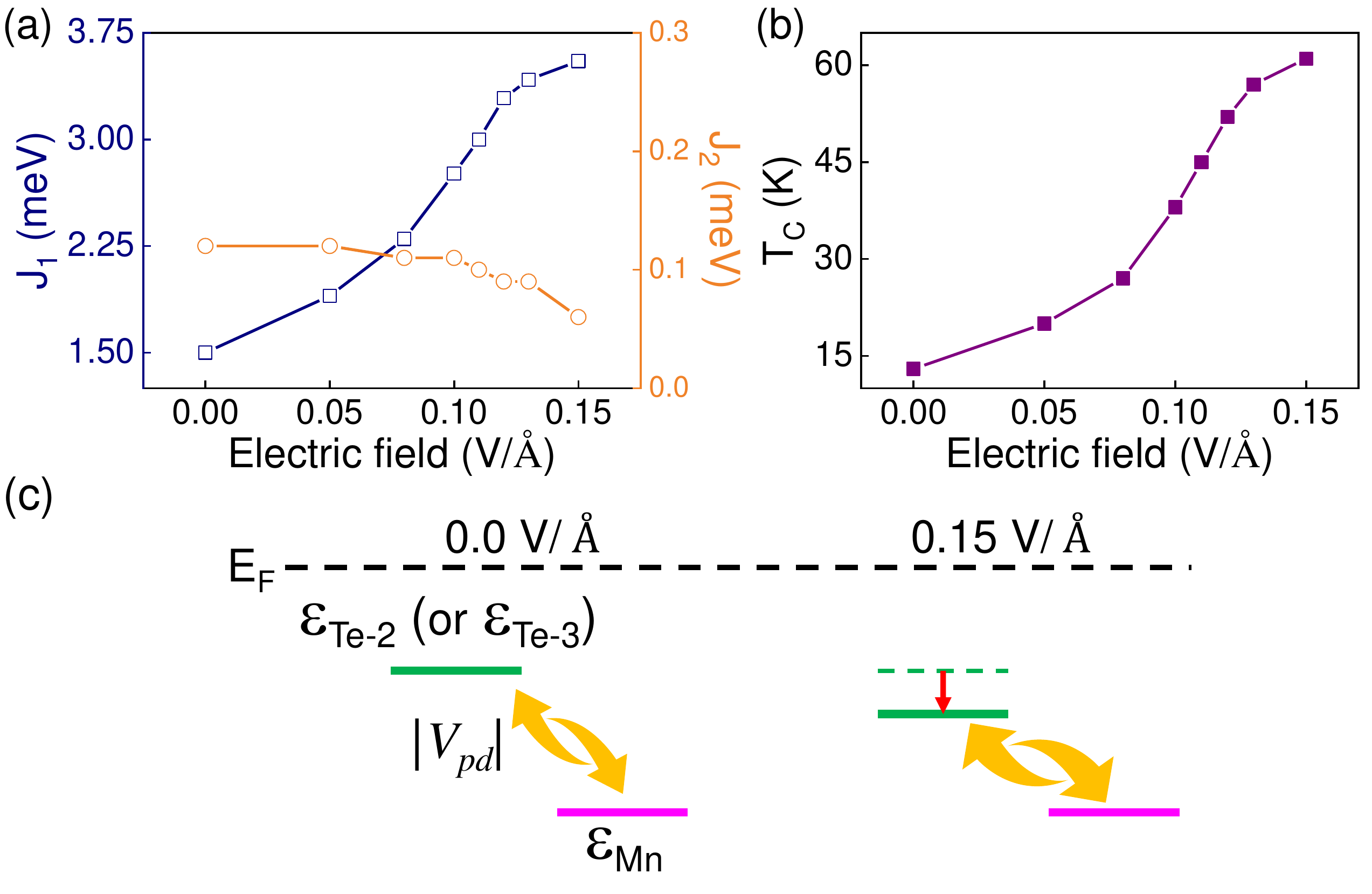}
\caption{Electric field enhanced Curie temperature. The evolution of the (a) exchange integrals and (b) Curie temperature with electric field. (c) The schematic diagram of the micromechanism of enhanced Curie temperature under electric field.}
\label{F-4}
\end{figure}

\begin{figure*}[!!!htbp]
\includegraphics[width=16cm]{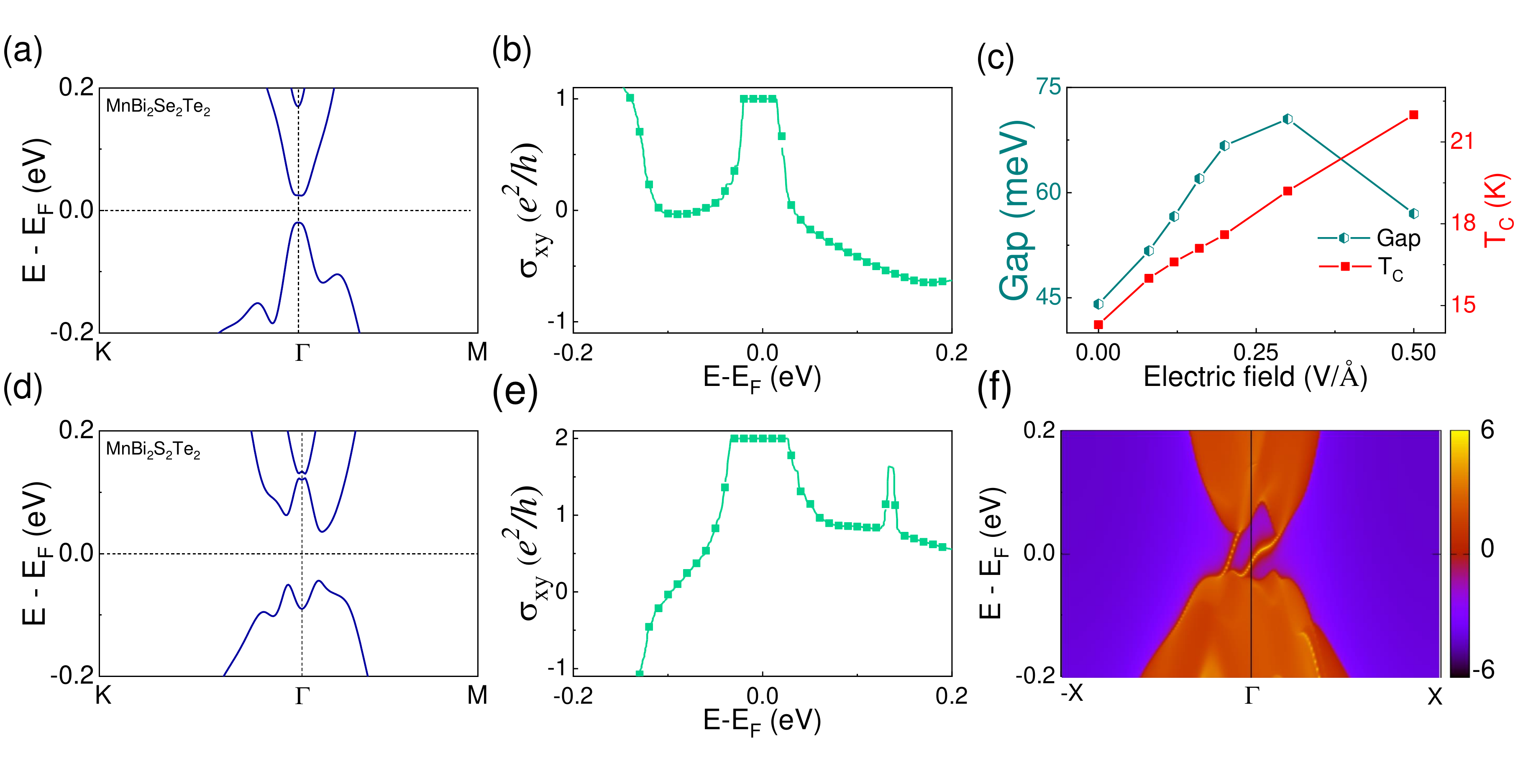}
\caption{QAHE in monolayers MnBi$_2$Se$_2$Te$_2$ and MnBi$_2$S$_2$Te$_2$. (a) Band structure, (b) AHC and (c) electric field dependent band gap and Curie temperature of monolayer MnBi$_2$Se$_2$Te$_2$. (d) Band structure, (e) AHC and (f) edge states of monolayer MnBi$_2$S$_2$Te$_2$.}
\label{F-5}
\end{figure*}

\subsection{Curie temperature under an electric field}
The exchange integral $J$ shares the similar dependence on the electric field, i.e. the exchange integral increases with the increase of E as shown in Fig.~\ref{F-4}(a). The increase of exchange integral $J$ can be interpreted by the superexchange interaction~\cite{Goodenough1955,Anderson1959,Kanamori1960}, where the FM coupling is expected since the Mn-Te-Mn bond angle is close to 90$^\circ$. The indirect FM coupling between Mn atoms is proportional to the direct AFM coupling between neighboring Mn and Te atoms. The magnitude of this direct AFM coupling can be roughly estimated as $J_{pd} = |V_{pd}|^2/|E_p-E_d|$, where $|V_{pd}|$ is the hopping matrix element between $p$ orbitals of Te and $d$ orbitals of Mn, and $|E_p-E_d|$ is the energy difference between them. By using maximally localized Wannier orbital projections, the dominant hopping matrix elements $|V_{pd}|$ and their corresponding energy differences $|E_p-E_d|$ can be obtained for monolayer MnBi$_2$Te$_4$ without E and with E=0.15V/\AA, respectively, as listed in Table~S2. The enhanced direct AFM coupling with increasing the electric field comes from the enhanced $p$-$d$ hybridization owing to the increase of hopping and the decrease of energy difference between $p$ orbitals of Te and $d$ orbitals of Mn. It can be understood from that the energy levels of Mn atoms are relatively deep in MnBi$_2$Te$_4$, and the electric field drives the energy levels of the upper and lower Te layers near Mn layer to move downward, resulting in the decrease of the energy level difference and the enhancement of the overlaps between $p$ orbitals of Te and $d$ orbitals of Mn atoms as schematically depicted in Fig.~\ref{F-4}(c).

The calculated Curie temperature based on the isotropic Heisenberg model with the single ion anisotropy (SIA) using Monte Carlo (MC) simulation was listed in Table~S1. It can be seen that for MnBi$_2$Te$_4$ with E=0.15 V/\AA\ the Curie temperature is even about 4.5 times larger than that without E as shown in Fig.~\ref{F-4}(b).

\section{Novel monolayers MnBi$_2$Se$_2$Te$_2$ and MnBi$_2$S$_2$Te$_2$ for QAHE}
Inspired by the above study, where an electric field can tune the energy band and induce a topological phase transition in MnBi$_2$Te$_4$ monolayer, we propose a 2D Janus structure MnBi$_2$Se$_2$Te$_2$ monolayer based on MnBi$_2$Te$_4$ monolayer. Janus crystals represent a class of 2D materials with different atomic species on their upper and lower facets. Janus MnBi$_2$Se$_2$Te$_2$ monolayer can be obtained by substituting the upper triple Te-Bi-Te layers of MnBi$_2$Te$_4$ monolayer with a triple Se-Bi-Se layers. The stability of MnBi$_2$S$_2$Te$_2$ monolayer was checked by its phonon spectra, molecular dynamic simulation and formation energy~\cite{SuplMat}. Because of different electronegativity of Se and Te, MnBi$_2$Se$_2$Te$_2$ monolayer possesses the spontaneous electric polarization of about 0.44 e\AA. MnBi$_2$Se$_2$Te$_2$ monolayer exhibits a ferromagnetic ground state with the in-plane magnetization, and its Curie temperature is estimated to be 14.3 K. Due to the small magnetic anisotropy, the magnetization direction can be tuned form in-plane to out-of-plane by a small magnetic field or strain. The band structure of MnBi$_2$Se$_2$Te$_2$ monolayer shows a full band gap of about 44 meV as shown in Fig.~\ref{F-5}(a). The calculation on anomalous Hall conductivity shows a plateau at the Fermi level, which corresponds to a Chern number $C=1$, indicating QAHE state as shown in Fig.~\ref{F-5}(b). As discussed above for MnBi$_2$Te$_4$ monolayer, when the system turns into the topological phase, the energy gap increases first and then decreases, and the Curie temperature keeps increasing with the increase of electric field. This is verified again in MnBi$_2$Se$_2$Te$_2$ monolayer, where the electric field enhances Curie temperature and changes the band gap as shown in Fig.~\ref{F-5}(c). 

For Janus material MnBi$_2$S$_2$Te$_2$ monolayer, its spontaneous electric polarization is about 0.73 e\AA, which is larger than that for MnBi$_2$Se$_2$Te$_2$ monolayer, thus a larger band gap and higher Curie temperature are expected in MnBi$_2$S$_2$Te$_2$ monolayer according to the above discussion. MnBi$_2$S$_2$Te$_2$ monolayer holds ferromagnetic ground state with out-of-plane magnetization, and its Curie temperature is estimated to be 30 K. MnBi$_2$S$_2$Te$_2$ monolayer has a global band gap of about 80 meV as shown in Fig.~\ref{F-5}(d). The topologically nontrivial band structure of MnBi$_2$S$_2$Te$_2$ monolayer is characterized by a nonzero Chern number $C=2$ with a quantized charge Hall plateau of 2$e^2/h$ and two gapless chiral edge states connecting the valence and conduction bands as shown in Figs.~\ref{F-5}(e) and (f), respectively.

To check the influence of the Hubbard $U$ on the QAHE, we calculate the anomalous Hall conductivity with $U$ = 3, 4 and 5 for monolayers MnBi$_2$Se$_2$Te$_2$ and MnBi$_2$S$_2$Te$_2$, respectively, as shown in Fig. S4. The results show that by varying the Hubbard $U$ from 3 to 5 eV, the QAHE states in both MnBi$_2$Se$_2$Te$_2$ and MnBi$_2$S$_2$Te$_2$ remain, but the band gap is slightly changed.

In monolayers one would expect a gap to open at an energy-scale set by the overlap of the edge mode wavefunctions on the top and bottom surfaces. The gaps opened by the mixture of the top and bottom surfaces for Janus MnBi$_2$Se$_2$Te$_2$ and MnBi$_2$S$_2$Te$_2$ are 118 and 121 meV, respectively, and the corresponding magnetic exchange energies of the above Janus materials are 140 and 161 meV. In both cases, the gaps are smaller than the magnetic exchange energy, supporting the QAHE states.

According to the calculated results of the phonon spectra, molecular dynamics simulation and the formation energy for Janus MnBi$_2$Se$_2$Te$_2$ and MnBi$_2$S$_2$Te$_2$, we found that they are both energetically and dynamically stable, and are feasible in experiments because of their exothermic reactions and much lower formation energy than that of MnBi$_2$Te$_4$~\cite{SuplMat}. Thus, one may expect to grow the bulk materials first and then exfoliate them. Alternatively, Janus MnBi$_2$Se$_2$Te$_2$ can be grown in the quintuple layers of Bi$_2$Te$_3$ and Bi$_2$Se$_3$ and a bilayer of MnTe with the molecular beam epitaxy method.

\section{Summary}
In this work, we systematically investigate the properties of monolayer MnBi$_2$Te$_4$ under electric field. It is found that the electric field can induce a topological phase transition from a topologically trivial state to a nontrivial state. It is also observed that the MAE and Tc of MnBi$_2$Te$_4$ monolayer are dramatically enhanced with the increase of electric field. The increased MAE mainly comes from the enhanced SOC due to the electric field. In terms of the super-exchange picture, the enhanced Tc can be understood from the enhanced $p$-$d$ hybridization and decreased energy difference between $p$ orbitals of Te atoms and $d$ orbitals of Mn atoms. In addition, we propose novel Janus monolayers MnBi$_2$Se$_2$Te$_2$ and MnBi$_2$S$_2$Te$_2$ with different spontaneous electric polarizations, which can realize the QAHE with Chern numbers $C$=1 and $C$=2, respectively. Our results will spur more studies on novel properties of MnBi$_2$Te$_4$ monolayer under an electric field, and also provide new candidates to realize QAHE in ferromagnetic semiconductors with electric polarization that would be particularly interesting in electrically controlled spintronic devices.

\section*{Acknowledgement}
Jing-Yang You and Xue-Juan Dong contributed equally to this work.

This work is supported in part by the National Key R\&D Program of China (Grant No. 2018YFA0305800), the Strategic Priority Research Program of the Chinese Academy of Sciences (Grant No. XDB28000000), the National Natural Science Foundation of China (Grant No.11834014), and Beijing Municipal Science and Technology Commission (Grant No. Z191100007219013). B.G. is also supported by the National Natural Science Foundation of China (Grants No. Y81Z01A1A9 and No. 12074378), the Chinese Academy of Sciences (Grants No. Y929013EA2 and No. E0EG4301X2), the University of Chinese Academy of Sciences (Grant No. 110200M208), the Strategic Priority Research Program of Chinese Academy of Sciences (Grant No. XDB33000000), and the Beijing Natural Science Foundation (Grant No. Z190011).

\end{document}